# Address-Specific Sustainable Accommodation Choice Through Real-World Data Integration

Peter J. Bentley[*], Rajat Mathur[†], Soo Ling Lim[‡], Sid Narang[§]
[*] *Dept. of Computer Science, University College London (UCL)*, London, United Kingdom
Email: p.bentley@cs.ucl.ac.uk
[†] *TheSqua.re*, London, United Kingdom
Email: rajat@thesqua.re
[‡] *Dept. of Computer Science, University College London (UCL)*, London, United Kingdom
Email: s.lim@cs.ucl.ac.uk
[§] *TheSqua.re*, London, United Kingdom
Email: sid.narang@thesqua.re

***Abstract*— Consumers wish to choose sustainable accommodation for their travels, and in the case of corporations, may be required to do so. Yet accommodation marketplaces provide no meaningful capability for sustainable choice: typically CO2 estimates are provided that are identical for all accommodation of the same type across an entire country. We propose a decision support system that enables real choice of sustainable accommodation. We develop a data-driven address-specific metric called EcoGrade, which integrates government approved datasets and uses interpolation where data is sparse. We validate the metric on 10,000 UK addresses in 10 cities, showing the match of our interpolations to reality is statistically significant. We show how the metric has been embedded into a decision support system for a global accommodation marketplace and tested by real users over several months with positive user feedback. In the EU, forty percent of final energy consumption is from buildings. We need to encourage all building owners to make their accommodation more efficient. The rental sector is one area where change can occur rapidly, as rented accommodation is renovated frequently. We anticipate our decision support system using EcoGrade will encourage this positive change.**

## I. Introduction

Consumers are now increasingly aware of the impact their behaviours have on our climate. But most find it difficult to know how to choose sustainable options [1]. The legislative drive towards global sustainability also means that many commercial sectors now need to quantify the sustainability of their products. Customers of appliances are provided with energy star ratings (USA) or energy labels (EU) that give easy-to-understand grades enabling choice [2]. If a customer wishes to buy a more energy-efficient washing machine, they can buy an A-rated machine instead of a B- or C-rated one. Such efficiency ratings are carefully allocated to specific machines through careful testing by regulatory bodies.

In the accommodation sector, travelers wish to choose sustainable options, but they are provided no valid information with which to make their choice. If a customer wishes to choose a more efficient apartment for their two-week visit they might be able to look at an approximation of its potential carbon footprint. But because of lack of data, the accommodation sector's approach to producing these estimates has been to use address-agnostic approximations that bear little resemblance to reality. The estimates are typically only specific to a country – certainly not to a region, or even city, meaning that no comparison can be made between options in the same country. A state-of-the-art net-zero building in London would be given the same (incorrect) value as a historic property with no insulation in Newcastle. Given the vast differences in sustainability caused by factors such as age, insulation, windows, main source of heating, appliances, etc., this has the effect that customers today can never know if their choice is truly sustainable nor not. But customers need to know, for many belong to corporations keen to reduce their emissions. And with millions of business and tourists renting apartments worldwide every week, enabling choice of sustainable options could have a positive impact on this growing industry.

Data is available to enable decision support in this area. In the accommodation sector, apartments and residential buildings should ideally have Energy Performance Certificates (EPCs) [3] in countries such as England, Scotland, Wales, France, Spain, Netherlands, or similar (e.g. CECB/GEAK [4] in countries such Switzerland). These certificates are the 'gold standard' when it comes to information about a specific property. Produced by a suitably qualified surveyor, they describe the property in terms of its energy efficiency (main heating source, lighting, insulation, window types), its location within a larger structure (e.g., ground floor apartment) and details of its construction (e.g., brick/wood, year of construction, floor area). They also estimate the likely energy usage and provide suggestions of how to improve the efficiency. The certificates are valid for up to ten years and may be required when selling or renting.

EPC datasets for a country may contain millions of records, sometimes multiple per property. They are ideal for our purposes of decision support. However, there are five fundamental problems with such data:

1. In the real world, a large proportion of buildings do not have any certification, or their certification is out of date (even when it may be mandatory for a property to be certified). We cannot look up the efficiency or energy usage of a property if it has no certificate.
2. In many countries, the data is not publicly available, falling under data privacy rules. For example, in

Switzerland there is excellent GEAK data, but it is only available (for a charge) for research purposes[1].

3. The data may be of poor quality with erroneous values found in the official database [5]. (In our own analysis we find that because of data entry errors, very inefficient buildings sometimes are mistakenly given a good rating in the EPC or vice versa.)

4. EPC calculations such as A-F ratings are country-dependent and change over time as government policies change.

5. In many countries there is no certification at all (e.g., most of Africa and Asia), or entirely different systems may be in use (e.g., the US Energy Information Administration datasets which provide city-specific approximations based on resident surveys) [6].

The aim of this work is to address this significant issue. We produce the first address-specific sustainability decision support system for property rental, validate it on data from ten UK cities and test it using a global apartment rental website: thesqua.re. Our solution makes use of multiple sources of data which are sanitized and consolidated to provide an easy-to-understand rating system for customers. We name the rating system EcoGrade and build a decision support system for renters and landlords.

The contributions of this work are as follows:

- The first ever address-specific rating for the accommodation sector, valid in multiple countries.
- Heuristic-based data integration and interpolation of multiple real-world data sources.
- Statistical validation of the rating using data comprising 10,000 UK properties and new analysis of accommodation sustainability.
- A real-world decision-support system using the rating, designed to provide choice for customers of rental accommodation.
- Testing of the system with real customers on a global accommodation rental website.

The rest of the paper is organized as follows: we provide a background in the next section, we describe the method in section III, validation and analysis in IV, real-world testing in V and we conclude in VI.

## II. Background

Carbon dioxide is the greenhouse gas that stays longest in our atmosphere and so in recent years the world has focused on using the weight of this gas (or equivalent) as a metric to quantify environmental impact – an idea first proposed in a 'successful, deceptive' marketing campaign from an oil company [7]. But the calculation of $CO_2$ relies on sufficient data which is rarely available, and we are now realizing that it may not encompass our impact effectively at all. The concept of 'carbon shadow' [8] provides a more nuanced idea: compare a frequent flier with someone who walks to work each day. A $CO_2$ metric tells you the former causes the worst environmental impact. But if the frequent flier is a climate scientist educating the world about dangers of climate change, and the walking commuter is an advertising agent for an oil company then who really causes the most harm long term? [8]

Work relating to sustainability and the rental market often focusses on profitability of energy efficiency retrofitting work. For example, [9] proposes RentalCal – a European rental housing framework to calculate profitability of energy efficiency retrofitting work. This work combines energy-demand modeling and retrofit option rankings with life-cycle cost analysis to quantify sustainable profits from retrofits [10].

EPCs are an excellent source of information about properties, but research shows that they must be used with care. For example, [11] examines the methods and input data used to generate EPCs for residential properties in the six largest European countries: UK, France, Germany, Spain, Italy and Poland, and found significant variation in the methods used to identify and assess energy consumption. We aim to bypass such variations by focusing on surveyed characteristics of dwellings; nevertheless, we recognize that any approach using EPC data will enable comparison *within* countries but not always *between* countries.

Where data is severely lacking, it is possible to use generative approaches to supplement datasets. In our previous work we develop an agent-based synthetic data generator (ASDG) to generate realistic data [12].

## III. Method

Understanding the sustainability of rental properties is a multi-factor calculation, which should make use of data regarding apartment efficiency, size, energy consumption, whether it makes use of green energy tariffs, and broader factors in the surrounding environment such as availability of green transport options. We use the principles of 'carbon shadows' to build as complete a picture as possible for users. Without considering all such factors, distortions can occur. For example, even should all factors relating to carbon be perfectly accounted for, a $CO_2$ figure alone would fail to encompass building efficiency: an inefficient poorly insulated building (or a building that consumes massive power) on a 'green energy tariff' might appear to have parity with a high-performance building, with the result that landlords would be given no impetus towards improving their building stock (or switching to energy efficient appliances). When considering the broader context of the stay, travelers may increase their carbon footprint via travel associated with their visit: business visitors typically need to travel regularly to workplaces; tourists travel for pleasure (sightseeing and activities). The calculation of sustainability should thus consider availability of green travel options such as electric scooters or public transport such as metro or underground trains. There are many APIs that provide live access to such facilities and which can be used to understand proximity to the rental address.

Our proposed solution is a novel data-driven decision support system that integrates multiple available data sources and intelligently adapts when data is unavailable or inaccurate. Our approach is designed to be applicable for all properties in all countries of the world. Here we focus on the UK countries England, Scotland and Wales to illustrate the methods used. (The method is also being applied to several European and North American countries; details of this ongoing work are extensive and beyond the scope of this paper.)

---

[1] https://www.geak.ch/der-geak/was-ist-der-geak/

TABLE I. EXAMPLES OF DATASETS USED FOR UK LISTINGS. (OTHER DATASETS ARE USED FOR LISTINGS IN OTHER COUNTRIES.)

| No. | Dataset / API | Provider | URL |
|---|---|---|---|
| 1 | Apartment listings | thesqua.re | proprietary data |
| 2 | Energy Performance of Buildings Data | Department for Levelling Up, Housing & Communities, UK government | https://epc.opendatacommunities.org |
| 3 | mysqua.re energy suppliers | mysqua.re | proprietary data |
| 4 | London bicycle share locations | Transport for London | https://api.tfl.gov.uk/BikePoint/ |
| 5 | London Underground station locations | Transport for London | https://tfl.gov.uk/info-for/open-data-users/api-documentation |
| 6 | Location of all all public transport access points | Department for Transport, UK government | https://www.data.gov.uk/dataset/ff93ffc1-6656-47d8-9155-85ea0b8f2251/national-public-transport-access-nodes-naptan |
| 7 | Location of bike, scooter, motor scooter, station, car points | Fluctuo | https://flow.fluctuo.com/api |
| 8 | Government conversion factors for company reporting of greenhouse gas emissions | Department for Energy Security and Net Zero and Department for Business, Energy & Industrial Strategy, UK government | https://www.gov.uk/government/collections/government-conversion-factors-for-company-reporting |
| 9 | Floor area to number of bedrooms | Mysqua.re | proprietary data |

### A. Industrial Partner

TheSqua.re is a digital corporate housing and luxury alternate accommodation marketplace with 200,000 furnished apartments managed by 2000 plus operators in 600 cities globally. They provide a large choice of alternate accommodations in major cities, using proprietary technology. In 2020, they launched MySqua.re, a private label brand that delivers a portfolio of homes in London offering city centre locations, currently live in more than ten neighbourhoods in London such as Fitzrovia, Mayfair, Kensington, Canary Wharf and City of London, operating more than 100 apartments. TheSqua.re Group (includes TheSqua.re and MySqua.re) aims to enable its customers (both building suppliers and renters) to transition to more sustainable practices. They provide data to enable this research and provide their platform for real-world testing.

Like most marketplaces, thesqua.re offers a clear and simple interface to search for available properties based on location, date and price. Our new decision support system integrates with this interface, offering renters several alternative ways to search and evaluate properties.

### B. Datasets

The main dataset is provided by thesqua.re, comprising rental apartment listings with addresses and information about number of bedrooms, bathrooms, plus photographs of the interior and sometimes exterior. Mysqua.re owned properties have guarantees for the accuracy of data. Smart meter readings are available for some. The remainder of the data comprises hundreds of thousands of property listings provided by external landlords. These listings provide their own problems, for in many cases a landlord may own multiple apartments and on receiving a booking request they may choose to make their own allocation. The result of this means that details such as internal photographs and reported apartment sizes may not always accurately represent the true dwelling that the customer will be given.

For each country we collate relevant data from the most reliable sources available. These datasets may be regional and so we need multiple sets to cover multiple cities. Table 1 lists examples of the datasets/APIs we use for UK, and the type of data available within them.

### C. Cleaning and Processing

Data in the apartment listings (Table 1, dataset No. 1) known to be unreliable, such as floor area is removed.

We use the EPC API (Table 1, No. 2) to download a matching EPC for every listing in dataset 1. Frequently there are no matching EPCs for a given listing in dataset 1. In this case we look for reports on neighboring properties – it is highly likely that neighboring apartments will share similar sustainability characteristics (they are often part of the same building), or that neighboring homes are similar in construction (they are often connected as terraced houses). To increase the chances that the neighboring homes are similar we look for EPCs in the same postcode. We consider *similar* properties only, where similarity is defined as having the same number of bedrooms. If there are insufficient similar properties we widen the search to look for properties in the same broader postcode (using the first 3 letters only).

Determining property similarity in terms of number of bedrooms is also non-trivial due to the lack of accurate data. EPCs do not contain number of bedrooms, only floor area in square metres. This is because owners may knock through walls or put up partition walls, decreasing or increasing the number of internal rooms without affecting the efficiency. But marketplace listings of floor areas are unreliable and cannot be used. So we transform floor areas reported in EPCs into probable number of rooms so that they can then be compared for similarity. We achieve the transformation using dataset 8, which is a look-up table of floor area ranges corresponding to studios, 1 bedroom, 2 bedroom, up to 5 bedroom properties, for all major cities in the UK. This dataset was prepared by mysqua.re using their proprietary industry data.

For (all) matching EPCs for the listing address, we clean the data, removing records that contain obvious data entry errors (e.g., a good rating but very large kwh/$m^2$ prediction, or implausibly large or small floor areas) and then download specific fields from the remaining records. (Where multiple records are found for a property we use the latest report; where duplicates are found with the same date we use the report with the worst rating.) The fields of interest are shown in Table II. Each field takes one of the following values: *very good, good, average, poor, very poor*, which we transform into numerical values: 1.0, 0.75, 0.5, 0.25, 0 respectively.

TABLE II. FIELDS DOWNLOADED FROM EPCS

| EPC Attribute | Meaning |
|---|---|
| HOT WATER ENERGY EFF | Efficiency of boiler used to heat water from the hot taps. |
| FLOOR ENERGY EFF | Insulation efficiency of the floor. |
| WINDOWS ENERGY EFF | Insulation efficiency of the windows. |
| WALLS ENERGY EFF | Insulation efficiency of the walls. |
| SHEATING ENERGY EFF | Efficiency of secondary heating, e.g., portable electric heaters. |
| ROOF ENERGY EFF | Efficiency of the roof. |
| MAINHEAT ENERGY EFF | Efficiency of the main heating, usually central heating. |
| MAINHEATC ENERGY EFF | Efficiency of main heating controller (e.g., is there central programmable control, thermostatic valves). |
| LIGHTING ENERGY EFF | Efficiency of the lighting. |

We also download the estimated energy consumption per metre squared. We do not use the estimated $CO_2$ figures in the report as the $CO_2$ estimation depends on energy production methods which vary over time as new powerplants come online; an older report will be out of date in its calculation. We also do not use the EPC final energy grade (a value of A to G, calculated using a standard assessment method that takes into account the individually measured factors) – while useful, this calculation changes over time as government reforms are introduced making the grades inconsistent. Such calculations also vary by country, adding yet more inconsistency. Instead, we focus on the observed characteristics of the property (heating, insulation, etc) that do not depend on changing government policies.

For APIs that provide location information on sustainable options such as bike points or stations (No.s 4, 5, 6, 7), we create our own dataset relevant to the apartment listings addresses. For every address we calculate its latitude $a_\phi$ and longitude $a_\lambda$. For every nearby sustainable travel option (electric bikes, scooters, station, etc) we find its latitude $s_\phi$ and longitude $s_\lambda$ and then find the Haversine distance between them (Eqn. 1):

$$h = 2r \times \sin^{-1}\left(\sin(s_\phi - a_\phi)^2 + \cos(a_\phi) \times \cos(s_\phi) \times \sin((s_\lambda - a_\lambda)/2)^2\right) \quad (1)$$

where we assume Earth's radius *r* to be 6371.

For APIs (e.g. no. 4) that provide current position of moving sustainable options such as electric scooters which may be left anywhere, we call the API multiple times during a week at different times of day and choose the nearest result.

We convert distances to time with the assumption that the average person can walk 5km/h and take the average of the distances to provide an overall figure representing the time to reach sustainable transport from apartment.

### D. *EcoGrade Calculation*

We integrate the multiple sanitized datasets to produce a combined set with valid values for each listed property; this enables the calculation of an *EcoGrade* score that will reflect its level of sustainability. The score will be a number between zero and five, so that it can be shown to users as a number of stars (similar to the star rating system familiar to anyone who has booked a hotel). EcoGrade uses four factors when available:

- Energy Consumption: if the average energy usage per metre squared is high (for example, air-conditioning and heated radiators on all the time) this means a high energy usage, costing the landlord more and increasing the carbon footprint. We use the predicted kwh/m$^2$ from the EPC(s), using a mean value if neighbouring property EPCs are being used (described above) because of a lack of EPC for this property. In some cases, we may be able to use direct smart meter readings from the property in question. The value is normalised and inverted as the score should be better for lower values.

- Energy Efficiency: a more efficient apartment (for example it is well-insulated) requires less power to achieve the same temperature. This costs the landlord less and reduces the carbon footprint. We measure efficiency based on detailed findings in the EPC report of the property (table 2). We take the mean value of all features retrieved. Again, if neighbouring property EPCs are used, we use the mean values for their features.

- Green Supplier: if the apartment has electricity provided by renewable energy (and does not use a gas boiler) then it will not generate $CO_2$ as energy is used. We make use of energy tariff details when available, awarding a maximum score when the tariff is 100% renewable energy. Where no data on suppliers is available, this factor is not used in the EcoGrade calculation.

- Green Transport: if there are many green options nearby then guests can travel to and from the apartment without generating so much $CO_2$. We make use of the average time calculated to green transport options (described above) and normalise the value, using the (generous) assumption that people will walk for an hour maximum.

The value for each factor is then transformed to be in the range 0 to 5 by a log function balanced to ensure a normal distribution of scores (i.e., best and worst ratings less common than average ratings).

An overall EcoGrade score is created by the average of all (available) factors. An additional $CO_2$ value is calculated using the predicted kwh/m$^2$ from the EPC, and making use of government provided conversion factors (Table 1, No. 8). Where neighbouring properties are used instead, a range of $CO_2$ values are provided using the best and worst matching neighbours' values, and an average value is provided of all matching neighbours' values.

### E. *Decision Support System*

While we can rate individual apartments using EcoGrade, it is important to provide this information in a manner that users will find intuitive and helpful. We achieve this in several ways: first, we enable apartments to be viewed in order of EcoGrade score (best to worst). We provide summary EcoGrade scores for all apartments given in search results. These are provided using a simple green logo with a number of 'leaves' underneath out of a maximum of five, Fig. 1.

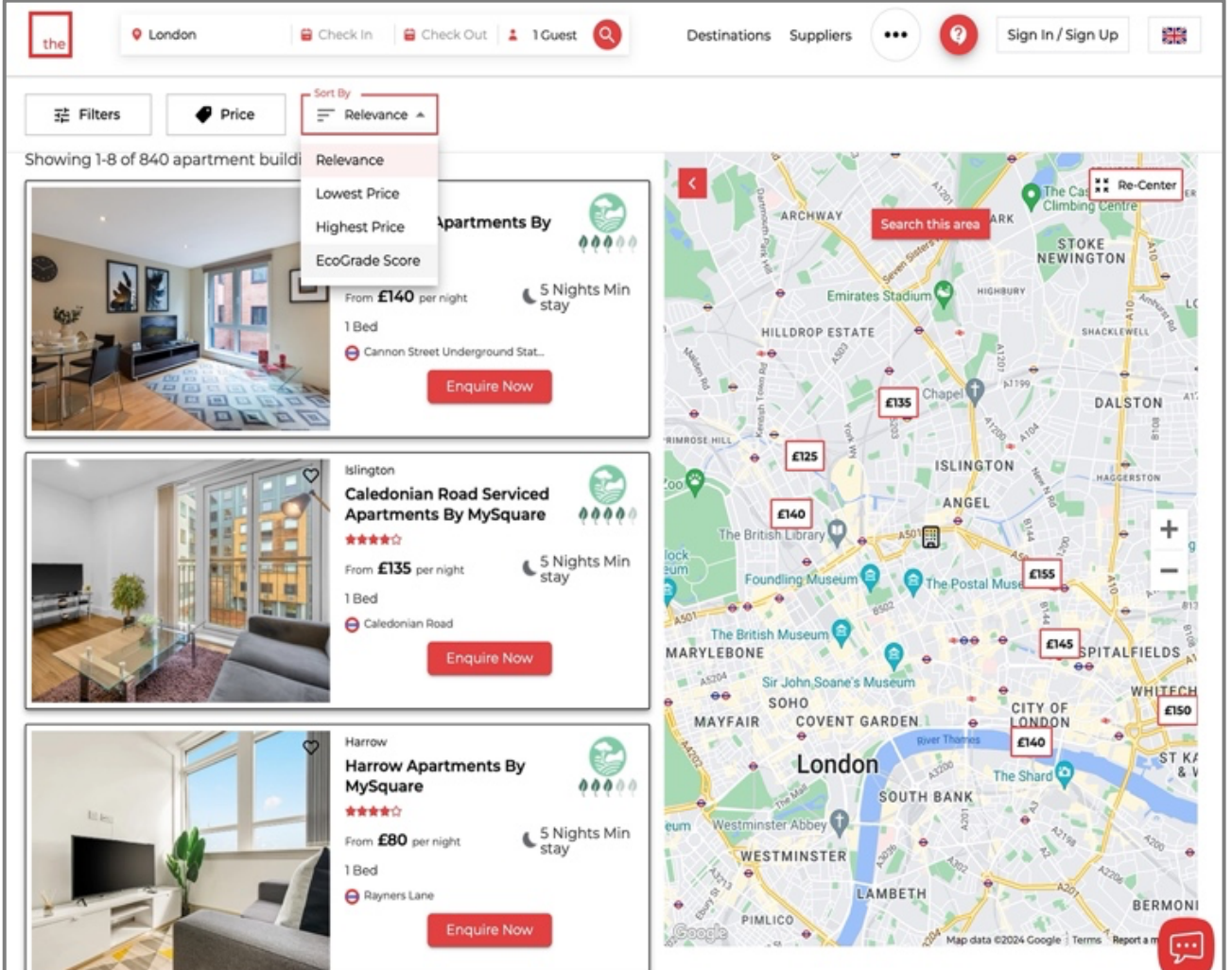

Fig. 1 Screenshot of search results showing green EcoGrade symbol and number of 'leaves' out of five.

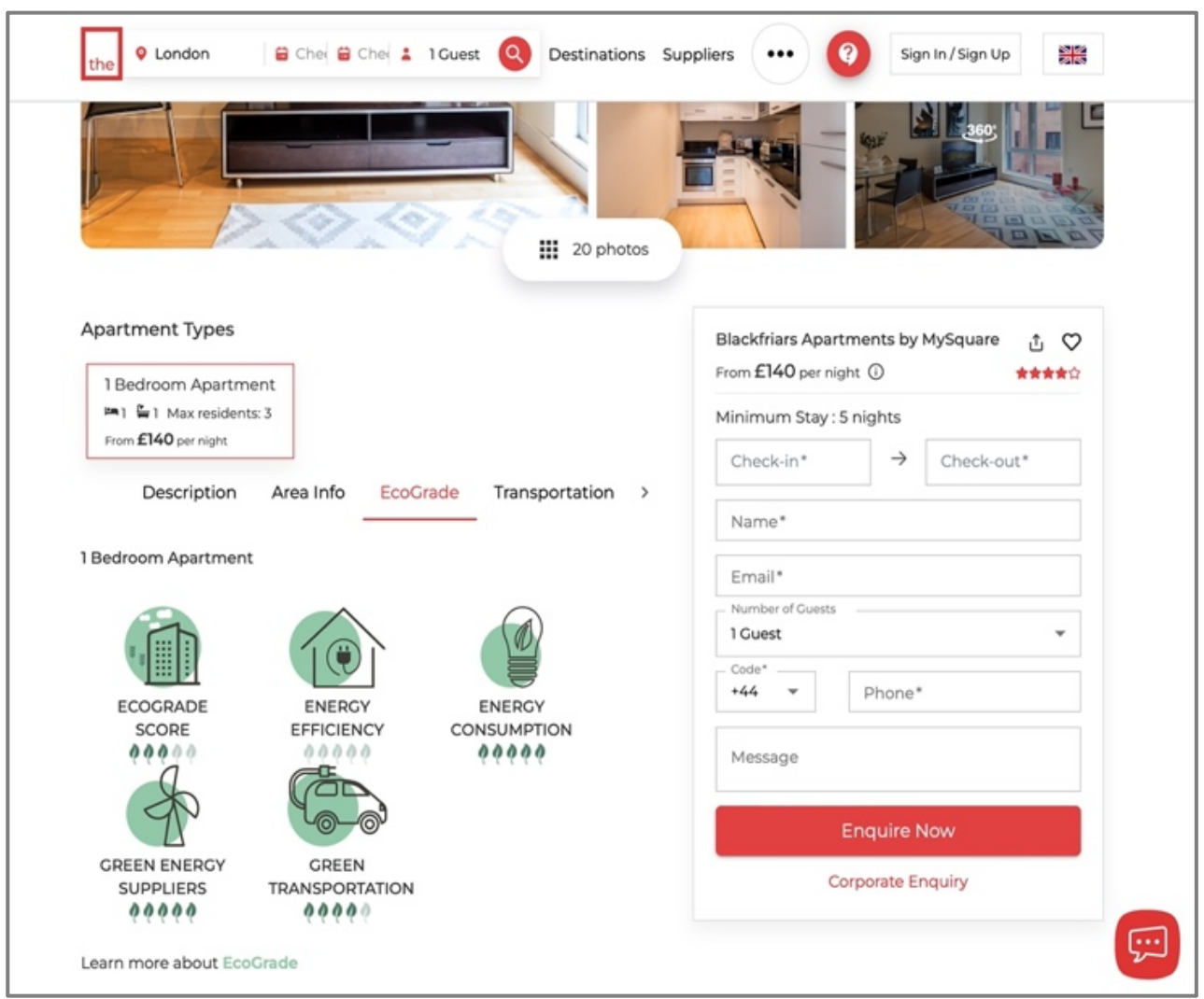

Fig. 2 The EcoGrade tab for an apartment shows the individual EcoGrade ratings.

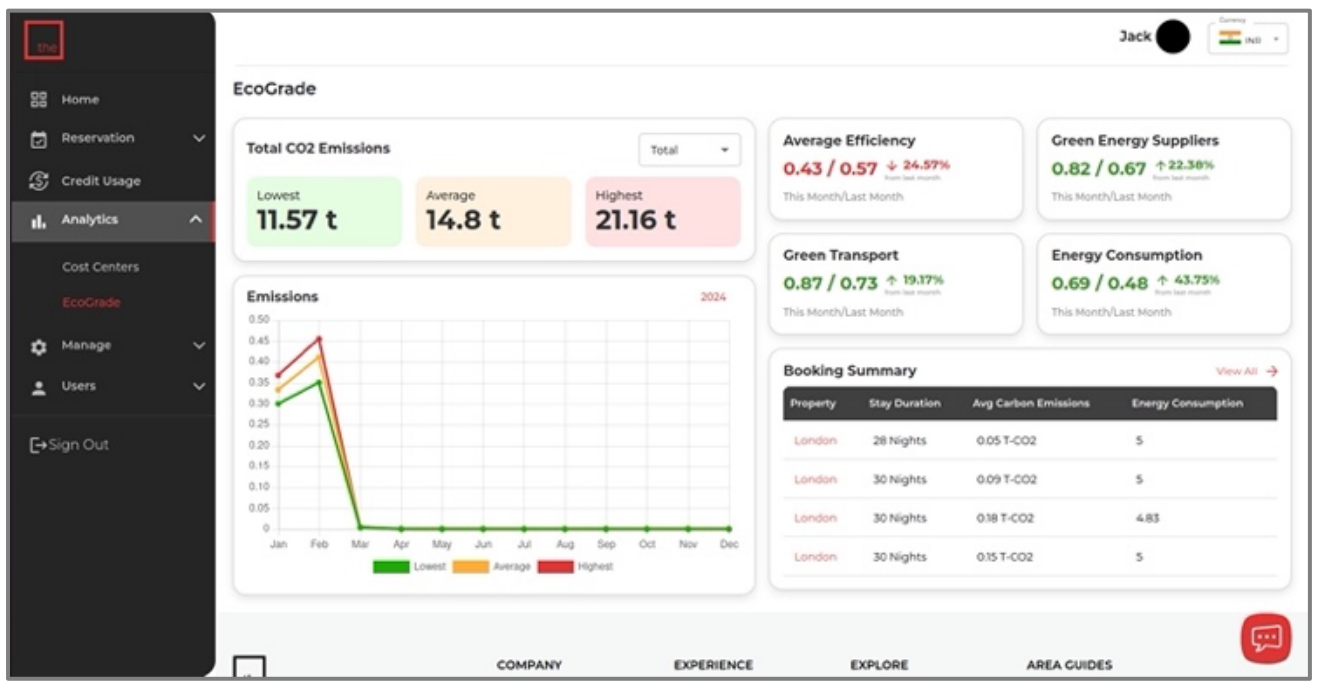

Fig. 3 Dashboard for corporate clients showing EcoGrade results of bookings over time. (Only Jan and Feb results available at time of writing; new data is added at the end of each month.)

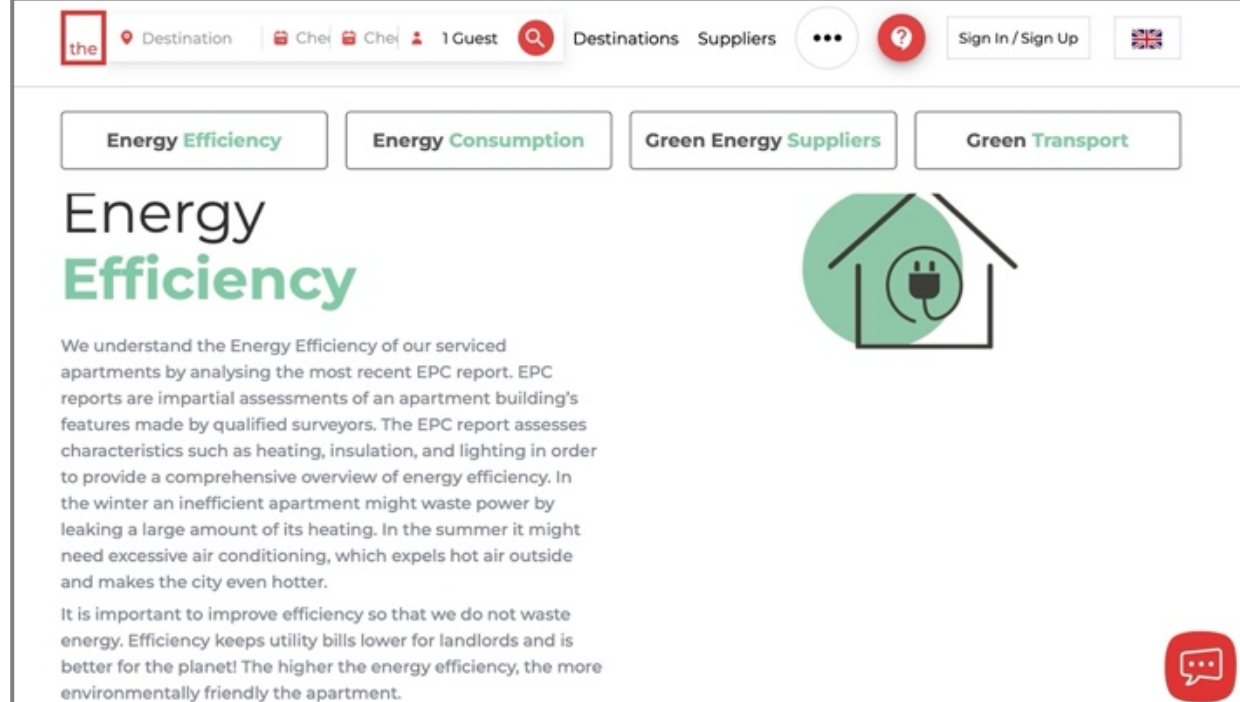

Fig. 4 EcoGrade explanation page for customers.

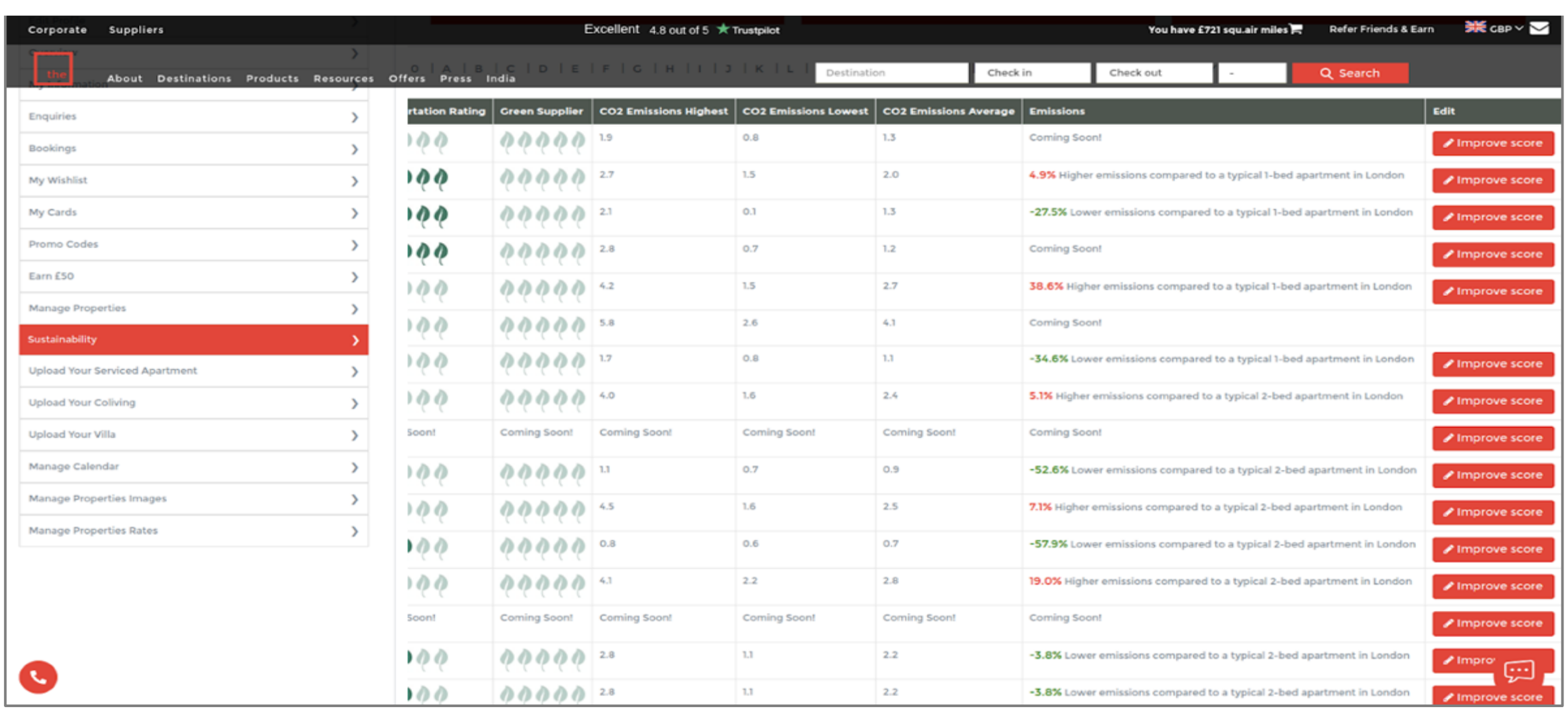

Fig. 5 Dashboard tool for building suppliers making use of Cohen's d percentage to provide $CO_2$ comparisons.

Clicking onto one apartment listing, and the user can click on a specific EcoGrade tab to see full details, Fig. 2. Corporate clients, who may make hundreds of bookings a year for their staff are provided with a dashboard to view their average EcoGrade booking scores, how each average EcoGrade feature is changing compared to the previous month and their $CO_2$ totals, Fig. 3. This enables users to understand individual factors and make their own assessments – e.g., if access to green transport is most important then the apartment shown in Fig. 3 may be a good choice. To help users understand and make informed choices, the website provides information about each factor accessed by clicking on "Learn more about EcoGrade", Fig. 4.We also provide a support tool for building suppliers, enabling them to understand the EcoGrade ratings and $CO_2$ for their listings, Fig. 5.We use Cohen's *d* percentage [13] to compare $CO_2$ ranges of each listing with average $CO_2$ ranges for that apartment type in that city, reporting e.g., 34.6% lower emissions compared to a typical 1-bedroom apartment in London, calculated as follows:

$$d = \frac{(a_\mu - c_\mu)}{\sqrt{((a_n - 1) \times a_\sigma{}^2 + (c_n - 1) \times c_\sigma{}^2)/(a_n + c_n - 2)}} \quad (2)$$

where:

$a_\mu$ is $CO_2$ emissions average for the apartment; $a_\sigma$ is $CO_2$ standard deviation for the apartment; $a_n$ is no. of samples considered for the calculation of the apartment EcoGrade;

$c_\mu$ is $CO_2$ emissions average for the city for a particular bed type; $c_\sigma$ is $CO_2$ standard deviation for the city for a particular bed type; $c_n$ is no. of samples considered for the calculation of EcoGrade for a particular bed type for the city.

Cohen's d percentage is a simple transform using the correlation coefficient:

$$d_p = 100 * \frac{d}{\sqrt{d^2 + 4}} \quad (3)$$

In addition to showing EcoGrade scores and how their apartments compare to average apartments of the same type in that city, the building suppliers can click on "Improve Score" and receive advice on how they can achieve higher EcoGrade scores (taken from advice provided in the EPCs). To improve data quality we also created an EcoGrade Programme for building suppliers. Members are encouraged to ensure EPCs are up to date, and details such as floor areas are correct for every listing, thus enabling accurate data without any need to interpolate data from neighboring properties.

## IV. Validation and Analysis

We first validate our EcoGrade scoring approach through analysis and comparison of scores generated across different cities of the UK. We generate 1000 random addresses for 10 major cities distributed across England, Scotland and Wales: Birmingham, Bristol, Cardiff, Edinburgh, Glasgow, London, Manchester, Milton Keynes, Newcastle, Nottingham. Addresses are generated within the UK postcode districts of each 'post town' for the city (generated by creating a random latitude and longitude value in the appropriate region and finding the nearest valid residential dwelling to that point). The data is hand-checked to remove anomalies, for example businesses on residential streets. Because we are simulating listings, we do not specifically pick dwellings with EPCs – this enables us to verify whether our approach of using neighboring properties to infer values for properties without EPCs provides statistically appropriate results. Where an address has no EPC we generate a random floor area in the range of neighbouring properties that do have EPCs.

We analyse the differences between mean EcoGrade scores for dwellings with EPCs and those where the scores were interpolated by using neighbouring properties, looking for differences. We compare results by city and by number of bedrooms. Fig. 6 illustrates the results (comparison by city), showing that orange (EcoGrade scores calculated directly from matching EPCs for the addresses) is a close match to the blue (EcoGrade scores calculated by interpolation from neighbours). We conducted a Two One-Sided Test (TOST) to determine if the mean scores of the two groups (G1: interpolated from neighbours' EPC and G2: directly calculated from EPC) are equivalent, with acceptance criterion set to 0.1. Both one-sided test *p*-values are significant (*p*-values = .000), indicating that there is statistically significant evidence of equivalence between the mean of the two groups. Additionally, the 95% confidence intervals for the mean scores were calculated as follows: G1 ((2.132, 2.1653) G2 (2.076, 2.114). These confidence intervals provide further insight into the precision of the estimated mean scores for each group. Consequently, we conclude that the means are equivalent, providing us with strong evidence that the EcoGrade interpolated from neighbours' EPC matches the EcoGrade directly calculated from EPC, overall.

The analysis also shows that results vary across cities, with Birmingham performing on average worse than London, for example. Results such as these enable another level of decision support as conference organizers or company facility managers could make use of such details to ensure a greater likelihood of sustainable accommodation being available.

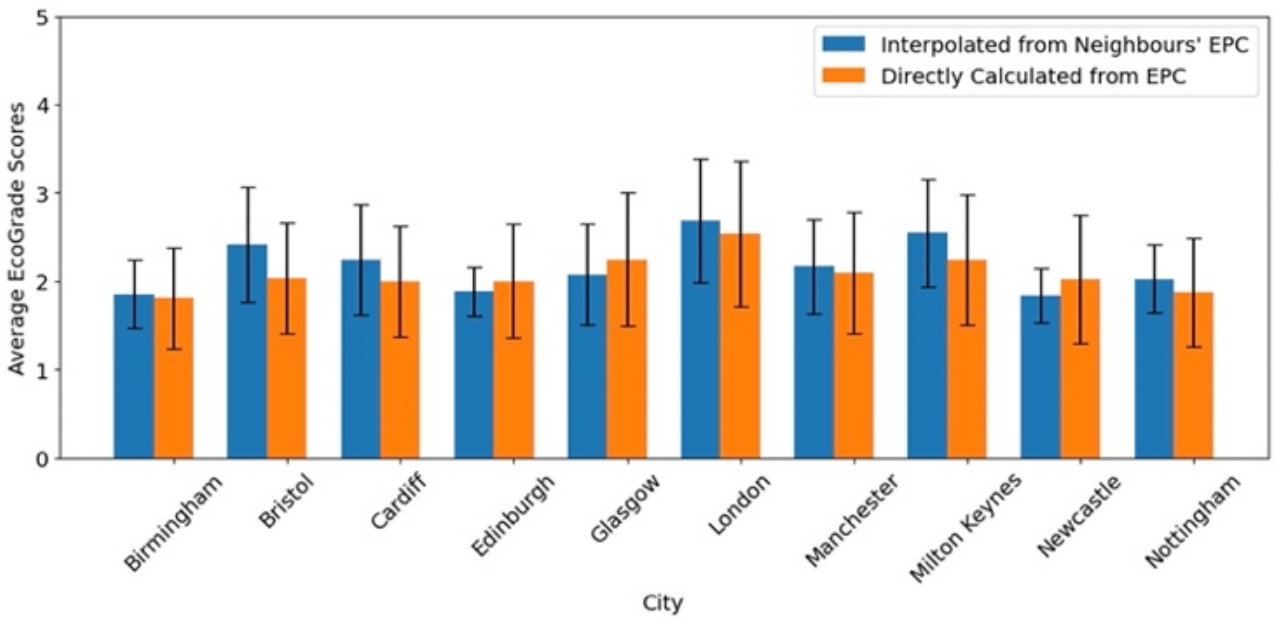

Fig. 6 City-wide comparison of EcoGrade scores comparing average scores generated directly from EPC matches (orange) with those generated by interpolating from neighboring properties' EPCs (blue).

Fig. 7 illustrates the distribution of EcoGrade scores for the UK, and illustrative example cities Birmingham and London, with the desired normal distribution curve showing lowest and highest scores are less common than average values, and that the distribution differs depending on the city, with Birmingham showing most dwellings have a score around 2, while London having greater diversity. This reflects reality well, as most cities in the UK have large quantities of older building stock with lower efficiency, but capital cities such as London attract considerable wealth which promotes regeneration and building of efficient dwellings, plus widespread sustainable transportation options. Fig. 8 (top left) shows the spread of EcoGrade scores for all ten cities, with London and newer city Milton Keynes (built in the 1970s and 80s) clearly providing more sustainable options compared to others. When analysing by bed types, it is apparent that more $CO_2$ is used for larger apartments as one would expect (Fig. 8 top right) however, there is little difference between studio and one bedroom apartments. The other charts show EcoGrade and its constituent features per bed type. It is clear that one to three bedroom apartments score comparatively well. The more bedrooms, the lower the energy consumption per metre squared. One, two and three bedroom dwellings have slightly better efficiency on average. Studio and one bedroom apartments tend to have better green transportation options (smaller apartments are often found more centrally in cities while large ones are further out). Taken together, this means that one bedroom apartments are likely to be an excellent choice for those wishing for a green option: the $CO_2$ footprint is no different from a studio, while it is more likely to be efficient and have sustainable transport options.

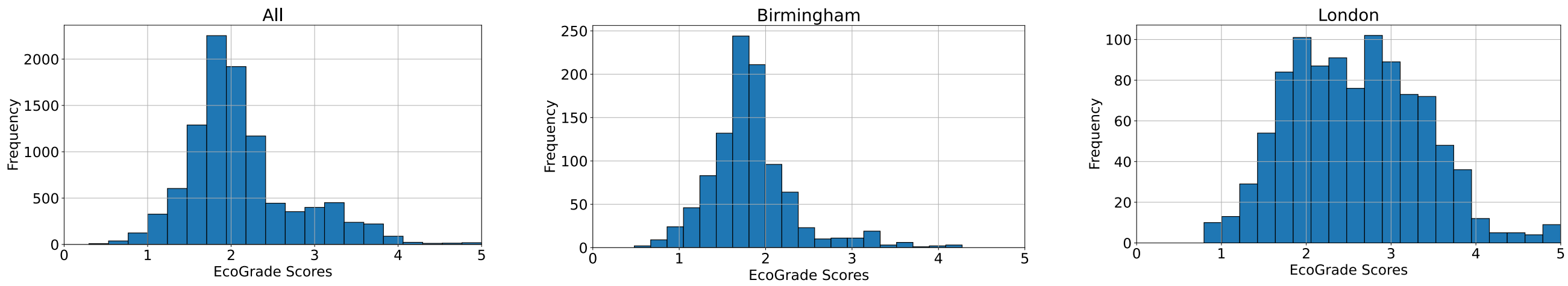

Fig. 7 Histograms of EcoGrade scores for all ten cities (left), Birmingham (middle) and London (right).

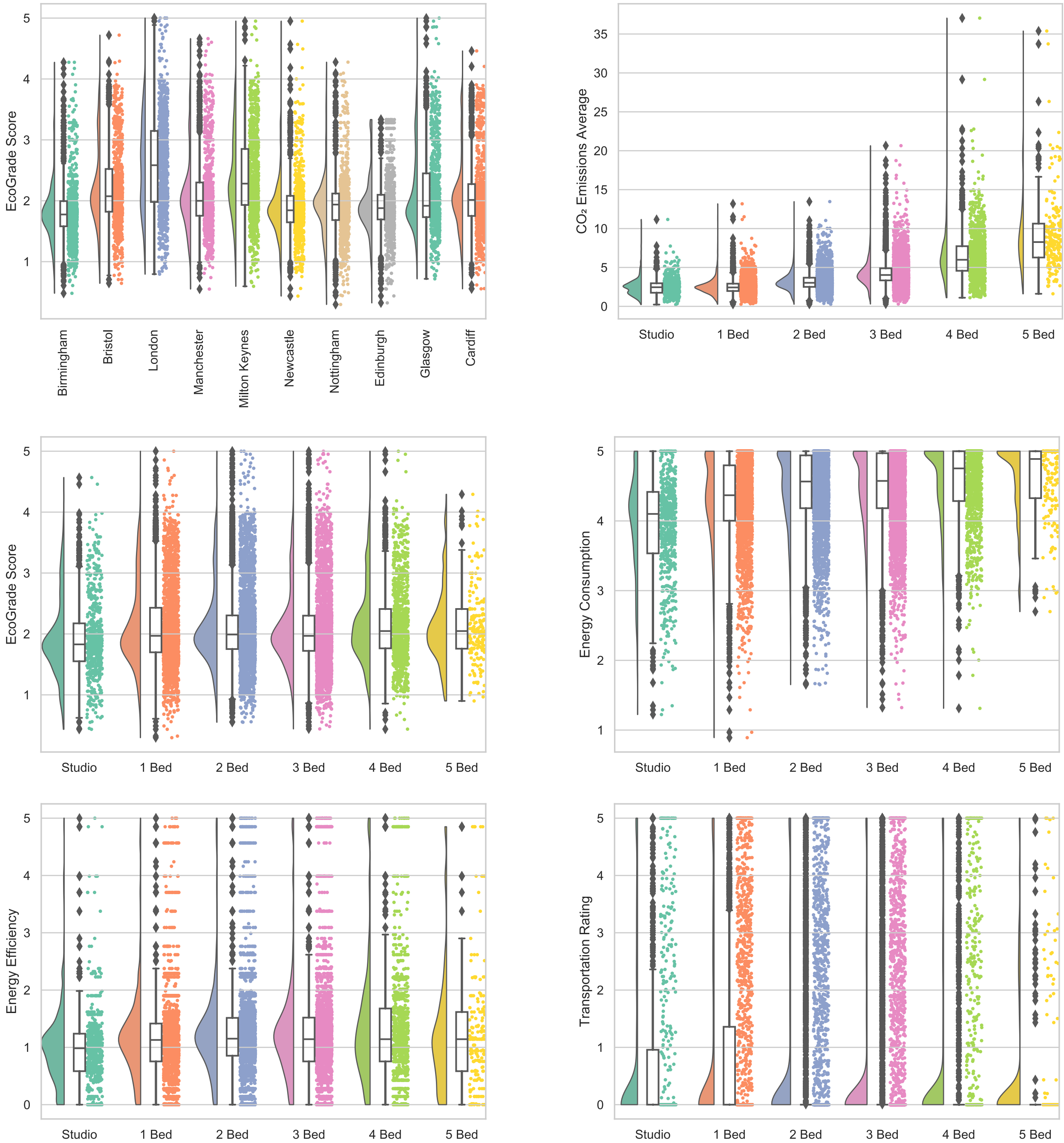

Fig. 8 Raincloud plots of EcoGrade per city (top left), $CO_2$ emissions in tons per annum per bed type (top right), and EcoGrade scores, and scores for individual EcoGrade features (bottom four). Green supplier not plotted as this data is not available for random UK properties.

## V. Real-World Testing

Having established the consistency of the EcoGrade metric, we now examine the effectiveness of the decision support system for users. To test this system, thesqua.re have trialled our EcoGrade search, information, and dashboard pages over several months, with thousands of real customers able to make use of it. In addition, ten building suppliers have actively engaged in the EcoGrade Programme, providing accurate address data on their hundreds of listings so that their EcoGrade scores can be calculated from EPCs accurately. In three months of usage the supplier EcoGrade dashboard was visited more than 400 hundred times, with more than 1600 hits to the EcoGrade explanation page. Building suppliers have reacted enthusiastically:

*"Excellent data offering for the corporates to consume and refer to it as a deciding factor for their accommodation options."*- Abhijit Sinha, Managing Director, EasyTrip India

*"Sounds brilliant"*- Coppergate Serviced Apartments

*"Well played. This is why we work with Green Tourism to understand our impact. And give a Green option for peeps*"- Scotty Hodson- Green Property Investor and Innovator, CEO at SILVA Executive Short Term Lets.

*"Easy to understand, this is brilliant, especially when everyone is so much more focused on our own impact on the environment"*- Sharon Baker, Senior Sales Manager

Comparisons with alternative approaches confirm why our decision support system using EcoGrade is a unique contribution. For example, a typical approach by another apartment marketplace uses methods that bear almost no connection to reality. Apartment $CO_2$ values are reported based on $CO_2$ reported for hotel rooms (also approximate), with that $CO_2$ value multiplied by the number of rooms in the apartment – the same values reported regardless of location. Hotels have very different energy usage compared to apartments and we find their estimated $CO_2$ values are typically more than twice ours (and those reported in EPCs) for all properties. For a few apartments, suppliers to the other marketplace self-reported apartment size and features such as air-conditioning and heating. A simple analysis shows the resulting "more accurate" $CO_2$ figures reported by this other marketplace are wildly unrealistic and do not match EPC figures, with different examples to be found of $CO_2$ an order of magnitude too high and too low. This appears typical of the industry today – $CO_2$ figures are not based on valid assumptions, they are not verified, analysed or checked sufficiently. In contrast, our approach uses a data-driven decision support system that uses the very best quality data available, generated by qualified EPC surveyors, and our assumptions are checked and verified through analysis.

## VI. Conclusion

Forty percent of final energy consumption in the EU was associated with buildings [11]. Where we choose to stay makes a real difference. Until now travelers were provided with nothing viable to help them make that sustainable choice. Here we described a decision support system for rental properties that provides a clear and data-driven choice to customers, and feedback to building owners. As part of the system we created the metric EcoGrade designed to encompass a holistic view of a stay: the efficiency of the property, the energy consumption, whether energy is supplied in a sustainable manner, and green transportation nearby. The metric is driven by integration of government-approved real-world datasets; inaccurate or missing data is rectified or interpolated. We validate the approach by generating 10,000 valid random residential addresses in 7 cities of England, 1 city in Wales, 2 cities in Scotland, checking our methodology and showing that EcoGrade can produce meaningful address-specific ratings for properties with and without EPCs. We then show how the decision support system has been used in the real world to enable choice of real apartments for a global accommodation marketplace, with positive user feedback.

The approach described here continues to be developed. Future work will investigate broadening the decision support system to more countries, making use of more datasets, and use AI to derive data from photographs, supplementing data further for data-poor regions of the world.